\documentclass[aps,prd,groupedaddress,showpacs,graphicx,nofootinbib]{revtex4}
\usepackage{amssymb,graphics,graphicx,color}
\def\lae{\;^{<}_{\sim} \;} 
\begin{document}

\title{Non-Gaussianity from two right-handed sneutrino curvaton decays}
%\author{Kazunori Kohri and Chia-Min Lin}
\author{Lu-Yun Lee$^1$}\email{d9522809@oz.nthu.edu.tw}
\author{Chia-Min Lin$^{12}$}\email{cmlin@phys.nthu.edu.tw}\affiliation{$^1$Department of Physics, National Tsing Hua University, Hsinchu, Taiwan 300}\affiliation{$^2$Institute of Physics, Academia Sinica, Taipei, Taiwan 115}

\date{Draft \today}

\begin{abstract}
In this paper, we consider the effect of two right-handed sneutrino curvaton decays and investigate the parameter space. We compare the difference of the result between single- and two-curvaton cases. We find one Yukawa coupling of the right-handed snetrinos can be as large as $\lambda \sim 0.1$ while the other one is much smaller. When the curvatons decay, we assume both of them subdominate the energy density of the universe. We find that, unlike a single curvaton case, here a small or negative, as well as a large $f_{NL}$ can be generated.
\end{abstract}
\maketitle

\section{Introduction}
Although the idea of inflation  \cite{Starobinsky:1980te, Sato:1980yn, Guth:1980zm} (for a review, see \cite{Lyth:2009zz}) started from solving the problems (like horizon and flatness problem) of the old hot big bang model, the key to distinguish different inflation models lies on primordial curvature perturbation $\zeta$ which provides the seeds for structure formation. Inflation could more or less generate some primordial curvature perturbation since we are living in a quantum world.
The primordial curvature perturbation generated from simple (single-field slow-roll) inflation models is adiabatic, almost scale invariant, and Gaussian. However, if inflaton is responsible for generating primordial curvature perturbation, the constraint of CMB normalization (i.e. $P_\zeta^{1/2} \simeq 5 \times 10^{-5}$) is so strong that we usually need to fine-tune the parameter(s) when building an inflation model. Furthermore, future experiments (like PLANCK satellite) may detect large non-Gaussianity and hence rule out simple scenarios of inflation. This will reveal for us a nontrivial way of generating primordial curvature perturbation and one of the promising ideas to generate large non-Gaussianity is through a curvaton \cite{Lyth:2001nq, Enqvist:2001zp, Moroi:2001ct}. Non-Gaussianity generated from curvaton scenario can be described by the nonlinear parameter $f_{NL}$, which takes the form
\begin{equation}
\zeta=\zeta_g+\frac{3}{5}f_{NL}\zeta^2_g+\cdots,
\end{equation}
where $\zeta_g$ denotes the Gaussian part of $\zeta$. Currently the upper bound of $f_{NL}$ is roughly given by ($2\sigma$) \cite{Komatsu:2010fb}
\begin{equation}
|f_{NL}| \lae 100.
\end{equation}
In the near future, the PLANCK satellite will reduce the bound to $|f_{NL}|<5$ if non-Gaussainity is not detected.

Curvaton is supposed to be light\footnote{In the context of cosmology, ``light" means mass smaller than Hubble parameter.} and subdominant during inflation. Because it is light, it can produce sizable quantum fluctuations which when stretched outside the horizon during inflation would become classical perturbations. Because it is subdominat, the perturbations should be regarded as isocurvature perturbation after inflation when the curvaton field starts to oscillate. The curvaton is supposed to decay after inflaton decay and at the same time transform its isocurvature perturbation to curvature perturbation. If the curvature perturbation of the universe is from the curvaton, it could liberate the constraint on inflation and lower the scale of inflation \cite{Dimopoulos:2002kt}. It would be interesting if we can identify a field from particle physics to be the curvaton and one of the candidate is a right-handed (RH) sneutrino in the framework of supersymmetry (SUSY). The idea of a right-handed sneutrino curvaton has already be mentioned in the original paper of curvaton \cite{Lyth:2001nq} and been considered in \cite{McDonald:2003xq, McDonald:2004by, Moroi:2002vx, Postma:2002et, Mazumdar:2004qv, Lin:2009yn, Lin:2009fk}. In this case, however, it is possible that we have more than one generation of the right-handed sneutrino and more than one curvaton. The calculation for two curvaton decay has been considered in \cite{Assadullahi:2007uw}, however, the parameters used in the paper is not of direct use for particle physicists to find possible candidate for the curvatons. In this paper we consider the decay width of the right-handed sneutrino curvaton in terms of the Yukawa couplings and masses. It turns out that there are six parameters in addition to the Hubble parameter. We then numerically scanning the parameter space in order to find solutions.

This paper is organized as followsing.
In section \ref{sec1}, we review the result of \cite{Lin:2009fk} in order to compare single- and two- curvaton decays in the succeeding sections. In section \ref{sec2}, we present the formalism we use for calculating two curvaton decay.  In section \ref{sec3}, we specify the parameter space and present our numerical results. Section \ref{sec4} is our conclusion. For completeness, there is also an Appendix section in the end of the paper where we summarize the equations adopted from \cite{Assadullahi:2007uw}.

\section{Single Right-Handed Sneutrino Curvaton}
\label{sec1}
It is well known that in standard model of particle physics, there is no good candidate to play the role of a curvaton. However, if we go beyond the standard model by imposing SUSY, there are lots of scalar fields. In this section we review the idea of a right-handed sneutrino which plays the role of the curvaton. We focus on the case that when the curvaton decays, it subdominates the energy density of the universe \cite{Lin:2009fk}, because large non-Gaussianity could be produced in this region of the parameters.

The superpotential of the RH neutrino is
given by
\begin{equation}
W_\nu=\lambda_\nu\Phi H_u L+\frac{m \Phi^2}{2},
\label{revision1}
\end{equation}
where $\Phi$ is the RH neutrino superfield, $H_u$ and $L$ are the
MSSM Higgs and lepton doublet superfields, and $m$ is the RH
neutrino mass. The canonical type-I seesaw mechanism gives the mass relation between light and heavy Majorana neutrino masses as
\begin{eqnarray}\label{numass}
m_{\nu} \sim \frac{\lambda^2_{\nu}v^2_{u}}{m}
\label{yukawac}
\end{eqnarray}
with $v_{u} \sim 10^2 \mbox{GeV} $ denoting the vacuum expectation value of $H_{u}$.
The mass squared differences revealed from the neutrino oscillation data, $\Delta m^2_{12} = 7.59^{+0.19}_{-0.21}\times10^{-5} {\rm eV^2}$ and $|\Delta m^2_{32}| = 2.43\pm0.13\times10^{-3} {\rm eV^2}$~\cite{pdg}, indicates that if we consider three generations of RH sneutrinos, the lightest left-handed neutrino mass can be very small while the mass of the other two generations are fixed due to the oscillation data. This means in principle we can have a Yukawa coupling arbitrarily small ($m_3 \sim 0$) but the other two Yukawa couplings should result in the neutrino masses around $m_{1,2} \sim 0.1\mbox{eV}$ (with a mass difference $\Delta m_{12} \sim 10^{-2}\mbox{ eV}$, namely, ``inverted ordering''). On the other hand, we can also have ``normal ording'' which implies  $m_1 \sim 0$, $m_2 \sim 0.01\mbox{ eV}$, and $m_3 \sim 0.1\mbox{ eV}$. From Eq.~(\ref{yukawac}) we can see that for the case $m_\nu \sim 0.1\mbox{ eV}$, if we have $m \sim 10^{-6}M_P$ ($m \sim 10^{-8}M_P$), we need $\lambda_\nu \sim 10^{-1}$ ($\lambda_\nu \sim 10^{-2}$).

The potential of the RH sneutrino $\sigma$ can be expressed as
\begin{equation}
V(\sigma)=\frac{1}{2}m^2\sigma^2.
\end{equation}
For simplicity, here we do not consider the Hubble-induced mass term. It is possible that the mechanism which suppresses the Hubble-induced mass term for the inflaton also suppresses that for the curvaton. For example, this is the case for D-term hybrid inflation \cite{Binetruy:1996xj}.
The decay rate for the RH sneutrino is
\begin{equation}
\Gamma=\frac{\lambda^2_\nu}{4\pi}m.
\end{equation}
The spectrum is given by
\begin{equation}
P^{1/2}_{\zeta_\sigma}=\frac{1}{3\pi}\Omega_{\sigma,D}\frac{H_\ast}{\sigma_\ast} \simeq 5 \times 10^{-5}
\label{eq1}
\end{equation}

If we assume that at the time $t_o$ of curvaton oscillation with
energy density $\rho_\sigma(t_o)=m^2\sigma^2_\ast/2$, the universe
is dominated by radiation (the decay products of inflaton) with
energy density $\rho_R(t_o)=3m^2 M_P^2$. At the time of
curvaton decay $t_D$, the energy density of the universe is given by
$\rho_R(t_D)=3\Gamma^2M_P^2=\rho_R(t_o)(a(t_o)/a(t_D))^4$. Therefore
$a(t_D)/a(t_o)=(m/\Gamma)^{1/2}$ and $\Omega_{\sigma, D}$ is given
by

\begin{equation}
\Omega_{\sigma,D} \equiv \left( \frac{\rho_\sigma}{\rho_{tot}} \right)_D = \frac{1}{6} \left(\frac{\sigma_\ast}{M_P}\right)^2 \left(\frac{m}{\Gamma}\right)^{1/2}=\frac{1}{6} \left(\frac{\sigma_\ast}{M_P}\right)^2 \frac{\sqrt{4 \pi}}{\lambda_\nu}.
\label{eq2}
\end{equation}
Throughout this paper, we always use a subscript ``$\ast$" to denote horizon exit. 
When $\Omega_{\sigma,D}$ is small, the nonlinear parameter is given by
\begin{equation}
f_{NL}=\frac{5}{4 \Omega_{\sigma,D}}.
\label{eq3}
\end{equation}
Note that here $f_{NL}$ can never be negative because we assume the curvaton is subdominant when it decays. As we will see in the following section, this is not true in the two-curvaton case.
By using Eqs.~(\ref{eq1}), (\ref{eq2}), and (\ref{eq3}), we can obtain 
\begin{equation}
\lambda_\nu M_P^2 =3.9 \times 10^3 \sigma_\ast H_\ast
\end{equation}
and
\begin{equation}
f_{NL}=(8.25 \times 10^3)\frac{H_\ast}{\sigma_\ast}.
\end{equation}
It is interesting  to note that here the spectrum does not give constriant on the mass of the right-handed sneutrino. However, as we can see in the next section, the results does depend on the masses when we consider the two curvatons case.

\section{Two Right-Handed Sneutrino Curvatons}
\label{sec2}

In our setup, we consider three generations of right-handed sneutrinos for type-I seesaw model and for simplicity we assume the heavist right-handed sneutrino (with a mass we denoted as $m_c$) does not play any role in cosmology, namely $m_c>H_\ast$.\footnote{If the gauge non-singlet scalar fields like a Higgs are light during inflation, the fields can establish an expectation value through quantum fluctuation. Through a Yukawa coupling, this gives an effective mass to the right-handed sneutrino field as can be seen in Eq.~(\ref{revision1}). The effect is most significant for the heavist right-handed snetrino due to the largest Yukawa coupling. Even if $m_c<H$, it is possible that the heavist right-handed sneutrino obtains an effective mass larger than the Hubble scale during inflation while other right-handed sneutrinos have masses lighter than the Hubble scale.} Here we consider the case that the lighter two right-handed sneutrinos are curvatons (we call them curvaton $a$ and curvaton $b$) and investigate the effects on the primordial curvature perturbation.  

The decay rates of the two curvatons are
\begin{equation}
\Gamma_a=\frac{\lambda_a^2}{4 \pi}m_{a}  \quad  \mbox{and}  \quad  \Gamma_b=\frac{\lambda_b^2}{4 \pi}m_{b}
\end{equation}
where $\lambda_a$ and $\lambda_b$ are the Yukawa couplings and $m_a$ and $m_b$ are their masses.
Quite generally, we consider 
\begin{equation}
H_\ast > m_a > m_b > \Gamma_a > \Gamma_b.
\label{eq9}
\end{equation}
We assume that when both of the curvaton decays, the energy density of the universe is dominated by radiation so that we can compare this with  single curvaton case. The similar calculation of the energy density ratio $\Omega$ as in Eq.~(\ref{eq2}) can be obtained for radiation $\gamma$, curvaton $a$, and curvaton $b$.
%\begin{equation}
%\frac{R_{ao}}{R_{aD}}=\left(\frac{\Gamma_a}{m_a}\right)^{\frac{1}{2}}
%\end{equation}
%\begin{equation}
%\frac{R_{bo}}{R_{aD}}=\left(\frac{\Gamma_a}{m_b}\right)^{\frac{1}{2}}
%\end{equation}
At curvaton $a$ decay (which is denoted by subscript ``1" throughout this paper),
\begin{eqnarray}
\Omega_{\gamma_0 1} &\simeq& 1, \nonumber \\
\Omega_{a1} &=& \frac{1}{6}\left(\frac{a_\ast}{M_P}\right)^2\left(\frac{m_a}{\Gamma_a}\right)^{1/2} = \frac{1}{6}\left(\frac{a_\ast}{M_P}\right)^2 \frac{\sqrt{4 \pi}}{\lambda_a}, \nonumber \\
\Omega_{b1} &=& \frac{1}{6}\left(\frac{b_\ast}{M_P}\right)^2\left(\frac{m_b}{\Gamma_a}\right)^{1/2} = \frac{1}{6}\left(\frac{b_\ast}{M_P}\right)^2 \left(\frac{4 \pi m_b}{\lambda_a^2 m_a}\right)^{1/2}.
\label{eq14}
\end{eqnarray}
where the subscript $\gamma_0$ denotes pre-existing radiation just before curvaton $a$ decay.
At curvaton $b$ decay (which is denoted by subscript ``2" throughout this paper),
\begin{eqnarray}
\Omega_{\gamma_1 2} &\simeq& 1 + \frac{1}{6}(\frac{a_{*}}{M_P})^2\frac{\sqrt{4\pi}}{\lambda_a}, \nonumber \\
\Omega_{b2}&=&\frac{1}{6}\left(\frac{b_\ast}{M_P}\right)\left(\frac{m_b}{\Gamma_b}\right)^{\frac{1}{2}}= \frac{1}{6}\left(\frac{b_\ast}{M_P}\right)^2 \frac{\sqrt{4 \pi}}{\lambda_b}.
\label{eq15}
\end{eqnarray}
where the subscript $\gamma_1$ denotes radiation just before
curvaton $b$ decay.

To linear order, the spectrum is given by \cite{Assadullahi:2007uw}
\begin{equation}
P_{\zeta(1)}=A^2P_{\zeta_{a(1)}}+B^2P_{\zeta_{b(1)}}
\end{equation}
where the parameters $A$ and $B$ can be found in the Appendix and
\begin{equation}
P^{1/2}_{\zeta_{a(1)}}=\frac{1}{3 \pi}\frac{H_{\ast}}{a_\ast}  \;\;\; \mbox{and } \;\;\;P^{1/2}_{\zeta_{b(1)}}=\frac{1}{3 \pi}\frac{H_{\ast}}{b_\ast}.
\label{revision2}
\end{equation}
Here we use subscript ``(1)" to denote ``first order" and ``(2)" will be used to denote ``second order".
If we define $\beta \equiv a_\ast/b_\ast$, we obtain
\begin{equation}
P^{1/2}_{\zeta(1)}=\left[A^2+\beta^2 B^2\right]^{1/2}\frac{1}{3\pi}\frac{H_\ast}{a_\ast}.
\end{equation}
We can write the spectrum explicitly by inserting $A$ and $B$ to obtain
\begin{eqnarray}
P^{1/2}_{\zeta(1)}
 &=&P^{1/2}_{\zeta(1)}(H_{*},a_{*},b_{*},m_a,m_b,\lambda_a,\lambda_b)
\nonumber\\&=& \frac{H_{*}}{3\pi}\sqrt\frac{\left[12M_P^2\sqrt{\pi}\frac{a_{*}}{\lambda_a}+4\pi\frac{{a_{*}}^3}{\lambda_a^2}+3\pi\sqrt{\frac{m_b}{m_a}}(\frac{a_{*}}{\lambda_a})(\frac{{b_{*}}^2}{\lambda_a})
\right]^2+\left[12M_P^2\sqrt{\pi}(\frac{b_{*}}{\lambda_b})+\pi{b_{*}}\left[\frac{a_{*}^2}{\lambda_a}(\frac{1}{\lambda_a}\sqrt{\frac{m_b}{m_a}}+\frac{3}{\lambda_b})+3\sqrt{\frac{m_b}{m_a}}\frac{b_{*}^2}{\lambda_a\lambda_b}\right]
\right]^2}{\left[4M_P^2+\sqrt{\pi}(\frac{a_{*}^2}{\lambda_a}+\sqrt{\frac{m_b}{m_a}}\frac{b_{*}^2}{\lambda_a})
\right]^2\left[12M_P^2+\sqrt{\pi}(4\frac{a_{*}^2}{\lambda_a}+3\frac{b_{*}^2}{\lambda_b})\right]^2}. \nonumber \\
\end{eqnarray}
This is subjected to CMB normalization $P^{1/2}_{\zeta(1)} \sim 5 \times 10^{-5}$ at horizon exit.

To second order, the curvature perturbation is given by \cite{Assadullahi:2007uw}
\begin{equation}
\zeta \equiv \zeta_2=\zeta_{2(1)}+\frac{1}{2}\zeta_{2(2)}=\left[A\zeta_{a(1)}+B\zeta_{b(1)}\right]+\frac{1}{2}\left[C\zeta^2_{a(1)}+D\zeta^2_{b(1)}+E\zeta_{a(1)}\zeta_{b(1)}\right]
\label{revision3}
\end{equation}
where $\zeta_2$ is the total curvature perturbation after the second curvaton decay. As in Eq.~(\ref{revision2}),  we use subscript ``(1)" to denote ``first order" and ``(2)"  to denote ``second order". The first order part $\zeta_{2(1)}$ is Gaussian and the second order part $\zeta_{2(2)}$ is non-Gaussian.
The parameters $C$, $D$, $E$ can be found in the Appendix
and the nonlinear parameter is
\begin{equation}
f_{NL}=\frac{5}{6}\frac{CA^2+\frac{1}{2}\beta^2 EAB+\beta^4 D B^2}{(A^2+\beta^2 B^2)^2}.
\end{equation}
\section{Numerical Results}
\label{sec3}
\begin{figure}[h]
  \centering
\includegraphics[width=0.5\textwidth, angle=-90]{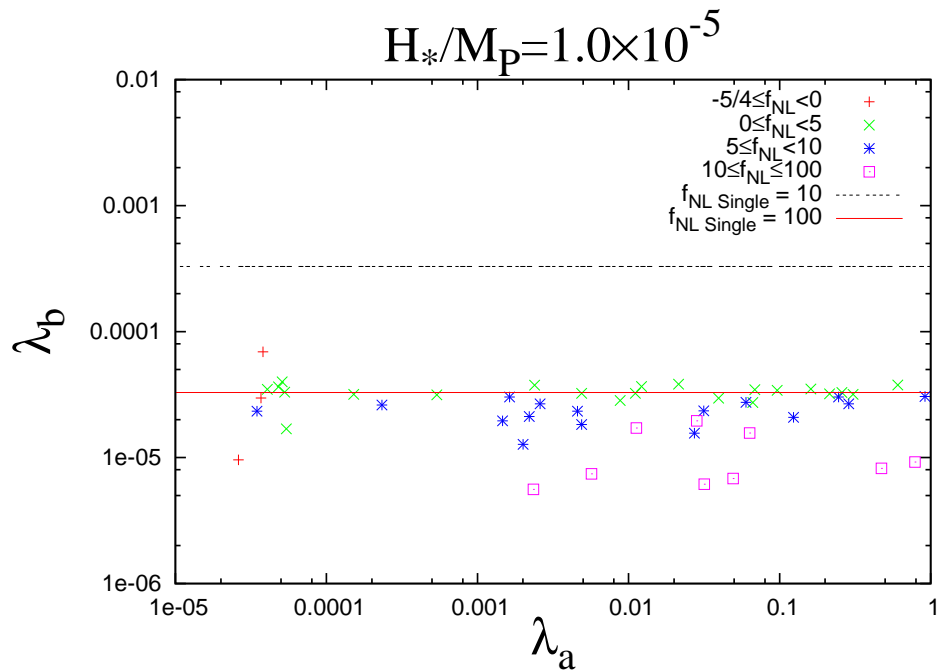}
  \caption{$\lambda_a$ versus $\lambda_b$. %Although we plot the Yukawa couplings here, each point represents a solution we found with all the six parameters fixed and this is true for all the plots in this paper.
  The points outside the range of single-curvaton bound represent the effect of considering two curvatons. As we can see here, one of the Yukawa couplings can be quite large.}
  \label{fig1}
\end{figure}

\begin{figure}[h]
  \centering
\includegraphics[width=0.5\textwidth, angle=-90]{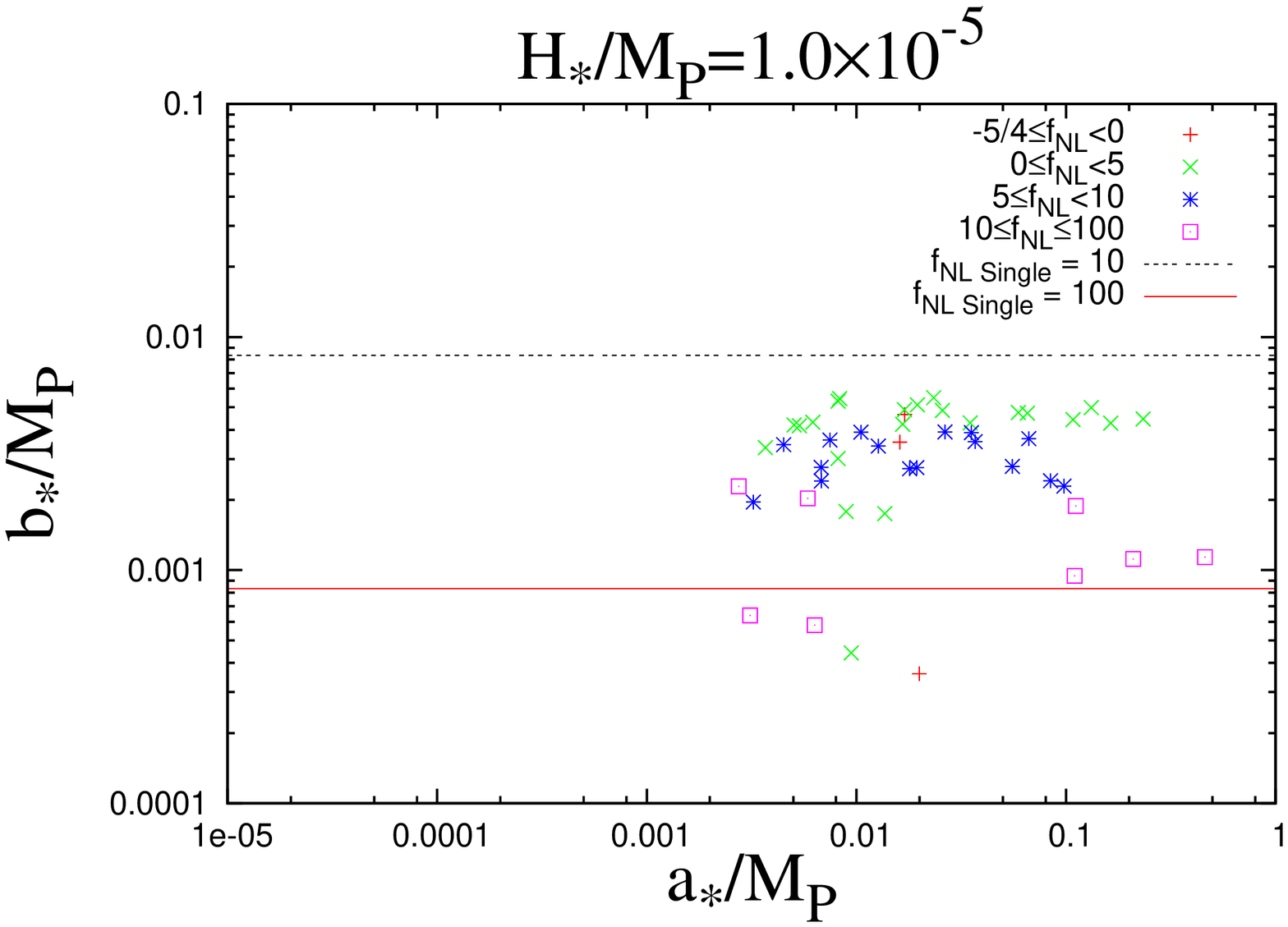}
  \caption{$a_\ast$ versus $b_\ast$.}
  \label{fig2}
\end{figure}

\begin{figure}[h]
  \centering
\includegraphics[width=0.5\textwidth, angle=-90]{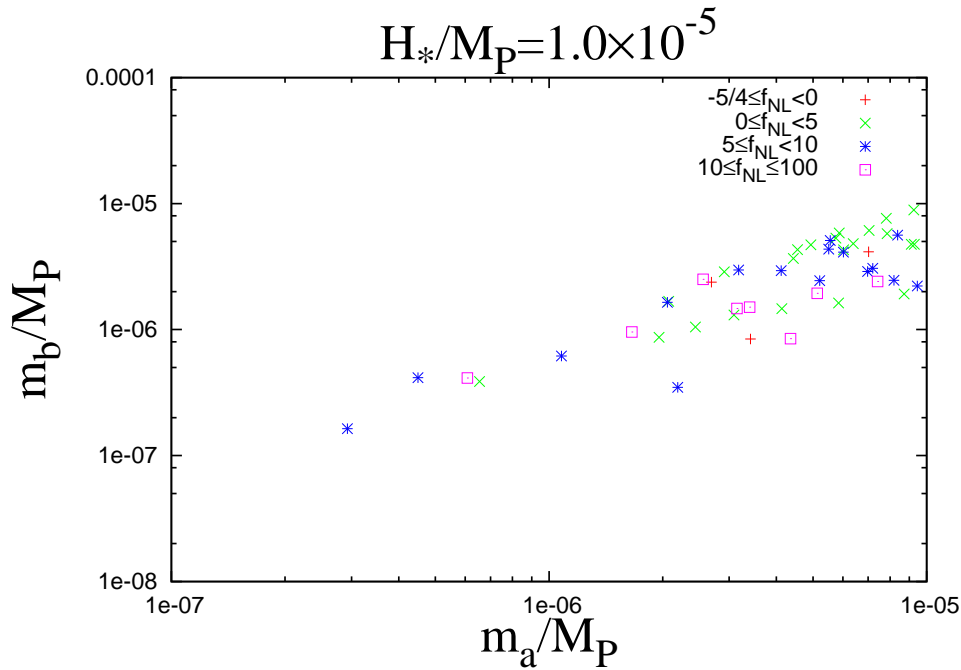}
  \caption{$m_a$ versus $m_b$. The apparent slope is due to the assumption that $m_a>m_b$.}
  \label{fig3}
\end{figure}

\begin{figure}[h]
  \centering
\includegraphics[width=0.5\textwidth, angle=-90]{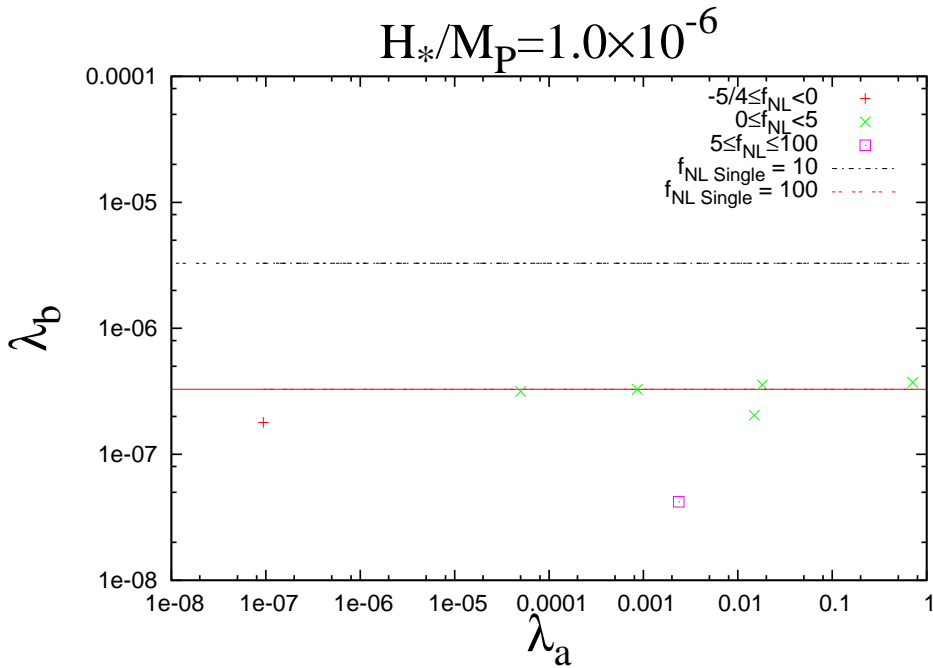}
  \caption{$\lambda_a$ versus $\lambda_b$. We found fewer points because it is more difficult to find solutions by our numerical computation for smaller $H_\ast$.}
  \label{fig4}
\end{figure}

\begin{figure}[h]
  \centering
\includegraphics[width=0.5\textwidth, angle=-90]{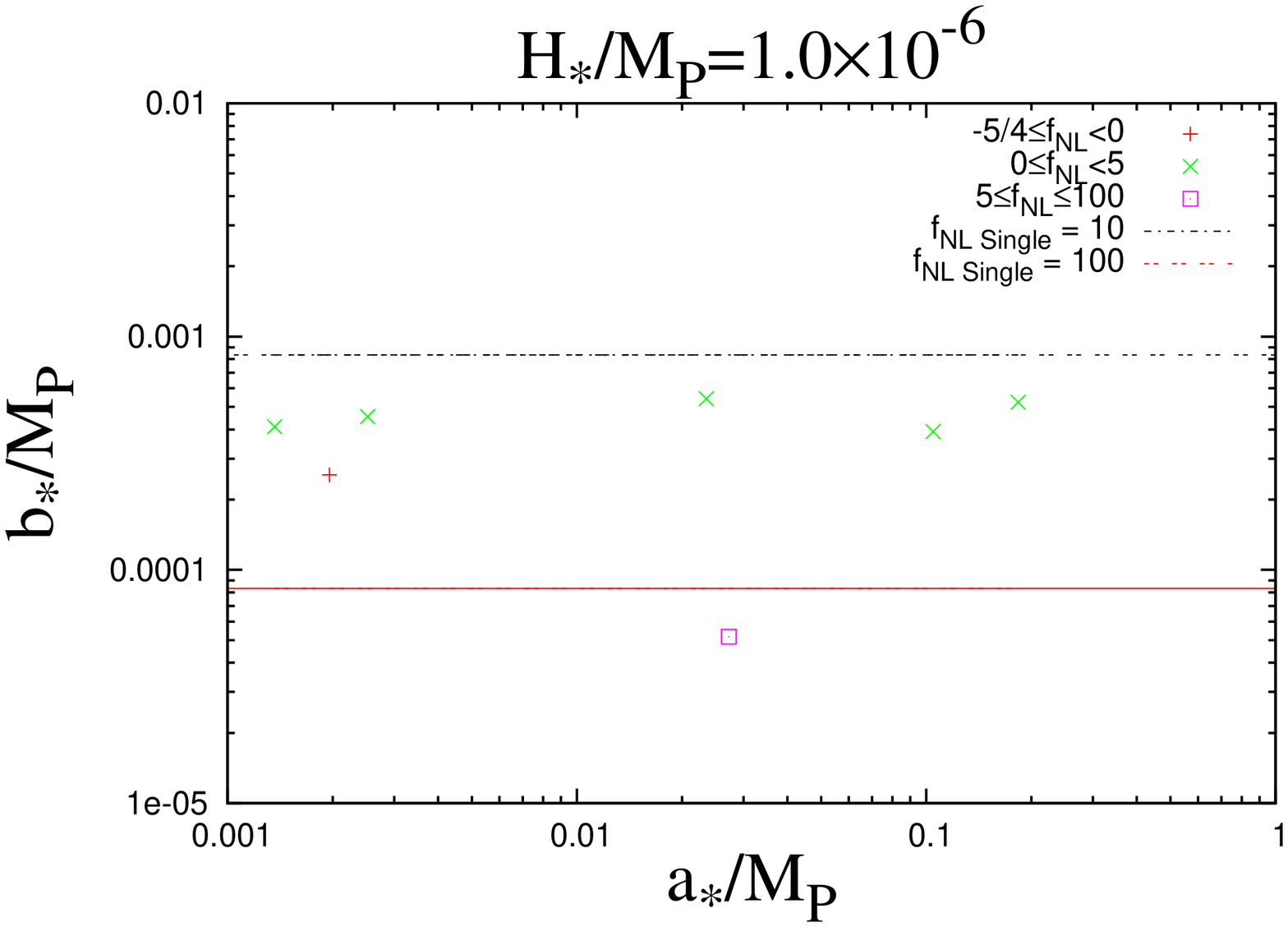}
  \caption{$a$ versus $b$.}
  \label{fig5}
\end{figure}

\begin{figure}[h]
  \centering
\includegraphics[width=0.5\textwidth, angle=-90]{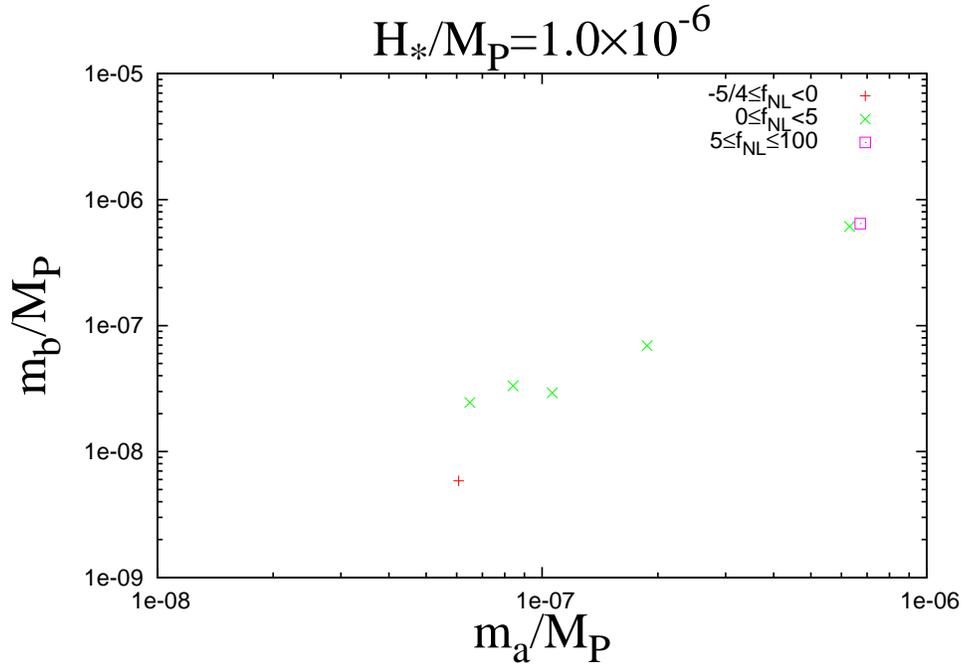}
  \caption{$m_a$ versus $m_b$.}
  \label{fig6}
\end{figure}

There are six parameters: $a_{*},b_{*},m_a,m_b,\lambda_a,\lambda_b$ in addition to the Hubble parameter $H_{*}$ and one constraint (CMB normalization). We tackle the problem numerically by scanning the parameter space. First of all, we assume the inflaton contribution to the curvature perturbation is small, this implies
\begin{eqnarray}
P^{\frac{1}{2}}_{\zeta_{inf}}=\frac{1}{2\sqrt{2}\pi}\frac{H_*}{\sqrt{\epsilon_H}M_P}<5\times10^{-5},
\end{eqnarray}
where $\epsilon_H \equiv -\dot{H}/H^2$. For a typical value of $\epsilon_H\sim0.01$ at horizon exit, we have the bound for Hubble parameter
\begin{eqnarray}
H_\ast<\sqrt{2}\pi\times10^{-5}M_P.
\end{eqnarray}
We will numerically find solutions by making plots for the two cases $H_\ast=10^{-5}M_P$ and $H_\ast=10^{-6}M_P$ in this paper.
The range of the curvaton field values at Hubble exit are chosen to be
\begin{equation}
H_\ast<a_\ast<M_P  \quad \mbox{and}  \quad H_\ast<b_\ast<M_P.
\end{equation}
The lower bound is from the requirement that the classical field value is larger than its fluctuation and the upper bound is from the requirement that the curvaton would not drive a second stage of inflation.
The masses of the curvatons are
\begin{equation}
10^{-15}M_P<m_a<H_\ast   \quad \mbox{and}  \quad 10^{-15}M_P<m_b<m_a.
\end{equation}
The lower bound is from the fact that the mass of the RH sneutrino should be larger than its soft mass which is assumed to be TeV scale and we also use the constraint from Eq.~(\ref{eq9}). The decay rate is also constrained by Eq.~(\ref{eq9}),
\begin{eqnarray}
\Gamma_a=\frac{\lambda_a^2m_a}{4\pi}<m_b,
\end{eqnarray}
therefore we choose
\begin{eqnarray}
10^{-10}<\lambda_a<\min\{\sqrt{\frac{m_b}{m_a}4\pi},1\}.
\end{eqnarray}
And again we require
\begin{eqnarray}
\Gamma_a=\frac{\lambda_a^2}{4\pi}m_a>\Gamma_b=\frac{\lambda_b^2}{4\pi}m_b,
\end{eqnarray}
therefore we choose
\begin{eqnarray}
10^{-10}<\lambda_b<\min\{\lambda_a\sqrt{\frac{m_a}{m_b}},1\}.
\end{eqnarray}
We plot our results in Figs.~\ref{fig1}-\ref{fig3} for $H_\ast=10^{-5}M_P$ and Figs.~\ref{fig4}-\ref{fig6} for $H_\ast=10^{-6}M_P$. For comparison, we also plot the results from the single-curvaton case. Naively one may think it is possible that one of the curvaton plays no role at all and the results would be dominated by a single curvaton. However, those plots show that both of the curvatons could play some role in generating curvature perturbation.
In all of the plots, each point represents a solution of all six parameters subjected to CMB normalization. We pick up a few points and list them on the Table~\ref{output}. In the plots $f_{NL}$ is an output after we imposing CMB normalization. We notice that even we assume the energy density of the curvatons is subdominant when they decays, we can still get small or negative (as well as large) $f_{NL}$. In addition, we also find that one of the Yukawa coupling can be as large as $\sim 0.1$ while the other coupling is relatively small. This is interesting because it is consistent with the neutrino oscillation data which we discussed below Eq.~(\ref{yukawac}) by the assumption that the curvatons are RH sneutrinos. For example, in the second row of  Table~\ref{output}, we would have the light neutrino masses correspond to the heavy neutrino masses $m_a$ and $m_b$ as $m_{\nu a} \sim m_1 \sim 0.1\mbox{ eV}$ and $m_{\nu b} \sim m_3 \sim 10^{-9}\mbox{ eV}$ respectively from  Eq.~(\ref{yukawac}). We can easily choose a $m_c>m_a$ with a larger Yukawa coupling to make $m_{\nu c} \sim m_2 \sim 0.11 \mbox{ eV}$  and make an inveted ordering of neutrino masses compatible with neutrino oscillation.  As another example, for the seventh row of Table~\ref{output}, $m_{\nu a} \sim m_2 \sim 0.01\mbox{ eV}$ can be obtained and along similar arguing we can obtain a normal ordering of neutrino masses. We would like to emphasis here that our goal is NOT showing that all of our parameter spaces are compatible with neutrino oscillation data, because our results can generically apply to other two-curvaton models with a similar decay rates. However, it is interesting enough if some of them do compatible with our assumption that two RH sneutrinos can play the role of curvatons.

\begin{center}
\begin{table}[h]
\begin{tabular}{|c|c|c|c|c|c|c|r|} \hline
%\multicolumn{8}{c}{\bf Table}
$\frac{H_{*}}{M_P}$ & $\frac{a_{*}}{M_P}$ & $\frac{b_{*}}{M_P}$ & $\frac{m_a}{M_P}$ & $\frac{m_b}{M_P}$ & $\lambda_a$ & $\lambda_b$ & $f_{NL}$   \\ \hline \hline                          	 
           $10^{-5}$ &$1.99\times10^{-2}$&$3.59\times10^{-4}$&$7.02\times10^{-6}$&$4.13\times10^{-6}$&$2.61\times10^{-5}$&$9.60\times10^{-6}$&$-0.95$ \\ %New Data H = 10^(-5)%
	  $10^{-5}$ &$2.33\times10^{-2}$&$5.49\times10^{-3}$&$4.13\times10^{-6}$&$1.46\times10^{-6}$&$6.05\times10^{-1}$&$3.76\times10^{-5}$&$2.88$ \\  %New Data H = 10^(-5)%
           $10^{-5}$ &$2.87\times10^{-1}$&$4.39\times10^{-3}$&$4.06\times10^{-6}$&$9.38\times10^{-7}$&$1.53\times10^{-1}$&$2.46\times10^{-5}$&$4.15$ \\ %New Data H = 10^(-5)%
	  $10^{-5}$ &$3.21\times10^{-1}$&$2.83\times10^{-3}$&$7.05\times10^{-6}$&$3.84\times10^{-6}$&$1.61\times10^{-1}$&$1.69\times10^{-5}$&$7.69$ \\  %New Data H = 10^(-5)%
	  $10^{-5}$ &$3.10\times10^{-3}$&$6.38\times10^{-4}$&$6.07\times10^{-7}$&$4.12\times10^{-7}$&$3.17\times10^{-2}$&$6.14\times10^{-6}$&$42.0$ \\  %Old Data H = 10^(-5)%
	  $10^{-6}$ &$1.95\times10^{-3}$&$2.55\times10^{-4}$&$6.07\times10^{-8}$&$5.85\times10^{-9}$&$9.39\times10^{-8}$&$1.79\times10^{-7}$&$-1.12$ \\ %New Data H = 10^(-6)%
	  $10^{-6}$ &$2.34\times10^{-2}$&$5.41\times10^{-4}$&$8.41\times10^{-8}$&$3.32\times10^{-8}$&$1.81\times10^{-2}$&$3.55\times10^{-7}$&$2.83$ \\ %New Data H = 10^(-6)%
	  $10^{-6}$ &$1.83\times10^{-1}$&$5.23\times10^{-4}$&$6.30\times10^{-7}$&$6.12\times10^{-7}$&$7.12\times10^{-1}$&$3.72\times10^{-7}$&$3.31$ \\  %Old Data H = 10^(-6)%
	  $10^{-6}$ &$1.04\times10^{-1}$&$3.91\times10^{-4}$&$1.87\times10^{-7}$&$6.92\times10^{-8}$&$1.49\times10^{-2}$&$2.04\times10^{-7}$&$4.82$ \\ %New Data H = 10^(-6)%
	  $10^{-6}$ &$2.72\times10^{-2}$&$5.17\times10^{-5}$&$6.72\times10^{-7}$&$6.44\times10^{-7}$&$2.36\times10^{-3}$&$4.20\times10^{-8}$&$52.0$ \\ \hline%New Data H = 10^(-6)%
\end{tabular}
\caption{\label{output} List of a few points of the numerical solutions.}
\end{table}
\end{center}

\section{Conclusion and Discussion}
\label{sec4}
We have explored the parameter space of generating primordial curvature perturbation via two right-handed sneutrino curvaton decays in the framework that there are three generations of RH sneutrinos. We compared the results with the single-curvaton case and found that an additional RH sneutrino curvaton could have some effects and cannot be neglected. Notably we may still get small or even negative $f_{NL}$ in the case that the energy density of the curvatons sub-dominates when they decay.  

As can be seen from Fig.~\ref{fig2} and Fig.~\ref{fig5} that the field value for the solutions may be not very small compared to the Planck mass (although it has to be smaller than the Planck mass in order to avoid driving a second stage of inflation). Therefore we have to suppress the nonrenormalizable terms in the superpotential. This can be achieved for example by judiciously assigning R-charge to the RH sneutrinos \cite{McDonald:2004by, Lin:2006xta}.

\appendix
\label{app}
\section{Notation}
The purpose of this appendix is to show how to obtain parameters $A$, $B$, $C$, $D$, and $E$ from Eqs.~(\ref{eq14}) and (\ref{eq15}).
The following equations and notations are from \cite{Assadullahi:2007uw}:\\

\noindent The energy density parameters ($\Omega_i=\rho_i / \rho_{tot}$):\\
$\Omega_{a1}$ the density of the first curvaton $a$ just before the first decay.\\
$\Omega_{b1}$ the density of the second curvaton $b$ at the first decay.\\
$\Omega_{b2}$ the density of the second curvaton $b$ just before the second decay.\\
$\Omega_{\gamma_01}$ the density of pre-existing radiation at the first (curvaton) decay.\\
$\Omega_{\gamma_11}$ the density of all radiation immediately after the first decay.\\
$\Omega_{\gamma_12}$ the density of all radiation just before the second decay.\\
$\Omega_{\gamma_22}$ the density of all radiation immediately after the second decay.\\

\noindent The full non-linear curvature perturbations:\\
$\zeta(=\zeta_2)$ the primordial perturbation after the second curvaton decay, but before the nucleosynthesis,\\
$\zeta_1$ the total perturbation at the first decay, \\
$\zeta_2$ the total perturbation at/after the second decay,\\
$\zeta_a$ the perturbation of the first curvaton $a$,\\
$\zeta_b$ the perturbation of the second curvaton $b$,\\
$\zeta_{\gamma_0}$ the pre-existing radiation perturbation,\\
$\zeta_{\gamma_1}$ the radiation perturbation after the second decay.\\

\noindent Subscript in parenthesis:\\
$\zeta_{(1)}$ the first order part of the primordial perturbation,\\
$\zeta_{(2)}$ the second order part of the primordial perturbation, \\
$\zeta_{1(1)}$ the first order part of the total perturbation at the first decay,\\
$\zeta_{1(2)}$ the second order part of the total perturbation at the first decay,\\
$\zeta_{2(1)}$ the first order part of the total perturbation at the second decay,\\
$\zeta_{2(2)}$ the second order part of the total perturbation at the second decay,\\
$\zeta_{a(1)}$ the first order part of the first-curvaton perturbation,\\
$\zeta_{a(2)}$ the second order part of the first-curvaton perturbation,\\
$\zeta_{b(1)}$ the first order part of the second-curvaton perturbation,\\
$\zeta_{b(2)}$ the second order part of the second-curvaton perturbation,\\
$\zeta_{\gamma_0(1)}$ the first order part of the pre-existing radiation perturbation,\\
$\zeta_{\gamma_0(2)}$ the second order part of the pre-existing radiation perturbation,\\
$\zeta_{\gamma_1(1)}$ the first order part of the radiation perturbation after the first decay,\\
$\zeta_{\gamma_1(2)}$ the second order part of the radiation perturbation after the first decay,\\
$\zeta_{\gamma_2(1)}$ the first order part of the radiation perturbation after the second decay,\\
$\zeta_{\gamma_2(2)}$ the second order part of the radiation perturbation after the second decay,\\

We have 
\begin{equation}
\zeta_{1(1)}=f_{\gamma_01}\zeta_{\gamma_0(1)}+f_{a1}\zeta_{a(1)}+f_{b1}\zeta_{b(1)}
\end{equation}
where

\begin{eqnarray}
f_{\gamma_0 1} &\equiv& \frac{4 \Omega_{\gamma_0 1}}{4 \Omega_{\gamma_0 1}+3 \Omega_{a1}+3 \Omega_{b1}}, \\
f_{a 1} &\equiv& \frac{3 \Omega_{a 1}}{4 \Omega_{\gamma_0 1}+3 \Omega_{a1}+3 \Omega_{b1}},\\
f_{b 1} &\equiv& \frac{3 \Omega_{b 1}}{4 \Omega_{\gamma_0 1}+3 \Omega_{a1}+3 \Omega_{b1}}.
\end{eqnarray}

After the first curvaton decays, but before the second curvaton decays, the curvature perturbation in the radiation at first order is given by
\begin{equation}
\zeta_{\gamma_1(1)}=R_1 \zeta_{1(1)}-(R_1-1)\zeta_{b(1)},
\label{a5}
\end{equation}
where

%R1%
\begin{eqnarray}
R_{1} &=& \frac{4-\Omega_{b1}}{4-4\Omega_{b1}}, \\
           &=& \frac{3+f_{a1}}{3(1-f_{b1})+f_{a1}}.
\end{eqnarray}

At the second decay, we have
\begin{equation}
\zeta_{2(1)}=f_{\gamma_12}\zeta_{\gamma_1(1)}+f_{b2}\zeta_{b(1)},
\label{a8}
\end{equation}
where

%f_r2,f_b2%
\begin{eqnarray}
f_{{\gamma_1}2} &\equiv& \frac{4\Omega_{{\gamma_1}2}}{4\Omega_{{\gamma_1}2}+3\Omega_{b2}}, \\
                    f_{b2} &\equiv& \frac{3\Omega_{b2}}{4\Omega_{{\gamma_1}2}+3\Omega_{b2}}
\end{eqnarray}

Finally, substitute Eq.~(\ref{a5}) into Eq.~(\ref{a8}), we obtain
\begin{equation}
\zeta_{2(1)}=R_1(1-f_{a1}-f_{b1})(1-f_{b2})\zeta_{\gamma_0 (1)}+r_a \zeta_{a(1)}+r_b \zeta_{b(1)},
\end{equation}
where

%r_{a},r_{b}%
\begin{eqnarray}
r_a &=& R_{1}f_{a1}(1-f_{b2}), \\
       &=& \frac{(1-f_{b2})(3+f_{a1})f_{a1}}{3(1-f_{b1})+f_{a1}} \\
r_b &=& 1-R_{1}(1-f_{b1})(1-f_{b2}), \\
       &=& \frac{(1-f_{b1})f_{b2}(3+f_{a1})+f_{b1}f_{a1}}{3(1-f_{b1})+f_{a1}}
\end{eqnarray}

Therefore if the pre-existing radiation perturbation vanishes, $\zeta_{\gamma_0(1)}=0$, at first order we can identify 

%A,B%
\begin{eqnarray}
A &=& r_{a}, \\
B &=& r_{b}.
\end{eqnarray}

At second order, we have
\begin{equation}
\zeta_{(2)} \equiv \zeta_{2(2)}=\tilde{C} \zeta^2_{a(1)}+ \tilde{D} \zeta^2_{b(1)}+E\zeta_{a(1)}\zeta_{b(1)}+F\zeta_{a(2)}+G\zeta_{b(2)},
\label{a18}
\end{equation}
where

%\tilde{C,D},E,F,G%
\begin{eqnarray}
\tilde{C} &=& -2R_{1}^2f_{a1}^2f_{b2}^2-R_{1}^2f_{a1}^2f_{b2}^3+7R_{1}^2f_{a1}^2f_{b2}-4R_{1}^2f_{a1}^2 \nonumber \\
                && +3R_{1}f_{a1}^2-f_{a1}^2-R_{1}f_{a1}^3-R_{1}f_{b1}f_{a1}^2+3R_{1}f_{a1}-3f_{b2}R_{1}f_{a1}^2 \nonumber \\
                && +f_{b2}f_{a1}^2+f_{b2}R_{1}f_{a1}^3+f_{b2}R_{1}f_{b1}f_{a1}^2-3f_{b2}R_{1}f_{a1} \\
\tilde{D} &=& -1+f_{b2}R_{1}f_{a1}f_{b1}^2-7R_{1}f_{b1}-f_{b1}^2+2f_{b1}+5R_{1}+f_{b2}-4R_{1}^2f_{b1}^2+8R_{1}^2f_{b1} \nonumber \\
                && +7f_{b2}R_{1}f_{b1}+7R_{1}^2f_{b1}^2f_{b2}-14R_{1}^2f_{b1}f_{b2}-2f_{b2}^2R_{1}^2f_{b1}^2+4f_{b2}^2R_{1}^2f_{b1} \nonumber \\
                && -f_{b2}^3R_{1}^2f_{b1}^2+2f_{b2}^3R_{1}^2f_{b1}-4R_{1}^2-5f_{b2}R_{1}+7f_{b2}R_{1}^2-2f_{b2}^2R_{1}^2-f_{b2}^3R_{1}^2 \nonumber \\
                && +3R_{1}f_{b1}^2-R_{1}f_{b1}^3+f_{b2}f_{b1}^2-2f_{b2}f_{b1}-R_{1}f_{a1}f_{b1}^2-3f_{b2}R_{1}f_{b1}^2+f_{b2}R_{1}f_{b1}^3 \\
           E &=& 2f_{a1}+2f_{b2}R_{1}f_{a1}f_{b1}^2-2R_{1}f_{a1}f_{b1}^2-10R_{1}f_{a1}-2R_{1}f_{b1}f_{a1}^2+10f_{b2}R_{1}f_{a1} \nonumber \\
               && +2f_{b2}R_{1}f_{b1}f_{a1}^2+14R_{1}^2f_{a1}f_{b2}f_{b1}-4f_{b2}^2R_{1}^2f_{a1}f_{b1}-2f_{b2}^3R_{1}^2f_{a1}f_{b1} \nonumber \\
               && -6f_{b2}R_{1}f_{a1}f_{b1}+8R_{1}^2f_{a1}-8R_{1}^2f_{a1}f_{b1}-14R_{1}^2f_{a1}f_{b2}+4f_{b2}^2R_{1}^2f_{a1} \nonumber \\
               && +2f_{b2}^3R_{1}^2f_{a1}+6R_{1}f_{a1}f_{b1}-2f_{a1}f_{b1}-2f_{b2}f_{a1}+2f_{b2}f_{a1}f_{b1} \\
           F &=& r_{a} = (1-f_{b2})f_{a1}R_{1} \\
           G &=& r_{b} = 1-R_{1}+f_{b1}R_{1}+f_{b2}R_{1}-f_{b2}f_{b1}R_{1}
\end{eqnarray}

Compare Eq.~(\ref{a18}) with Eq.~(\ref{revision3}), we conclude

%C,D%
\begin{eqnarray}
C &=& \tilde{C}-\frac{3}{2}F, \\
D &=& \tilde{D}-\frac{3}{2}G
\end{eqnarray}

The non-linearity parameter is:

%f_{NL}%
\begin{eqnarray}
f_{NL}=\frac{5}{6}
\frac
{
r_{a}^2(\tilde{C}-\frac{3}{2}r_{a})+\frac{1}{2}\beta^2Er_{a}r_{b}+\beta^4r_{b}^2(\tilde{D}-\frac{3}{2}r_{b})
}
{
{(r_{a}^2+\beta^2r_{b}^2)}^2
}
\end{eqnarray}

\acknowledgments
CML was supported by the NSC under grant No. NSC 99-2811-M-007-068.
We would like to thank Kingman Cheung and Chian-Shu Chen for discussion during the work.

\end{document}